\documentclass[12pt,a4paper]{article}
\usepackage{amssymb,amsmath,amscd,epsfig}
\title{{\bf Surface operators and magnetic degrees of freedom in  Yang-Mills
theories}}
\author{A. Di Giacomo\\University of Pisa
and INFN, Sezione di Pisa, \\Largo Pontecorvo 3, 56127, Pisa, Italy\\
V.I. Zakharov\footnote{Electronic address: vzakharov@itep.ru}\\INFN, Sezione di Pisa, Largo Pontecorvo 3, 56127, Pisa, Italy;\\ITEP, B. Cheremushkinskaya 25, Moscow, 117218, Russia}

\begin{document}
\maketitle

PACS:  	 12.38.Aw; 11.15.Ha

\abstract{ Magnetic degrees of freedom are manifested through
violations of the Bianchi identities and associated with singular
fields. Moreover, these singularities should not induce color
non-conservation. We argue that  the resolution of the constraint is that the
singular fields, or defects are Abelian in nature.
Recently proposed surface operators seem to represent a general solution
to this constraint and can serve as a prototype of
magnetic degrees of freedom. Some basic lattice
observations, such as the Abelian dominance of the confining fields, are
explained then as consequences of the original non-Abelian invariance.
Generically, the properties of the two-dimensional defects
associated with the surface operators are close to the predictions
of the dual models for the magnetic D2 branes.
\footnote{This paper is prepared for the volume dedicated to the 80th birthday
of Lev Borisovich Okun. The authors are acknowledging deep influence which
Lev's Okun style and works have had on them. One of the main themes of papers of L.B.
Okun is that it is only Nature, or experiment which can decide whether a phenomenon
is exotic or not, reality or an artefact. We would include nowadays lattice
data, or better to say lattice experimentation with field theory into
the means to discover new phenomena. Magnetic monopoles were introduced
first, as a pure theoretic construct more than 70 years ago by P.M.A Dirac
\cite{dirac}. Reality of descendants of the Dirac monopoles, magnetic degrees
of freedom in Yang-Mills theories was established, we believe, via lattice
experiment.  In these notes we are trying to make a step towards theoretical
interpretation of these magnetic degrees of freedom invoking for this
purpose the notion of surface operators \cite{gukov} which proved useful in such mathematical
discipline as number theory. We are amused by the idea that a kind of Abelian dominance
exists both in Yang-Mills theory and in the number theory and the same mathematical
means can be adapted to describe the two phenomena. We hope that the simplicity
of the basic means used will match Lev's Okun quest for clarity of the basic
concepts of physics.
}}

\section{Introduction}

It is commonly accepted that confinement is due to
condensation of magnetic degrees of freedom. Such a mechanism is
established, both theoretically and via lattice measurements in the
Abelian cases of a pure gauge field in the compact version of the
theory and $Z_2$ gauge theory, for an early review
see  \cite{savit}. In this case the magnetic degrees of freedom
are in fact the Dirac monopoles.

Moreover on the lattice one observes monopole trajectories
and has to use  the polymer approach to field theory, see, e.g. \cite{ambjorn},
to describe the monopole properties in the language of the quantum geometry.
In the geometrical language the confining
field configurations are
identified  as infinite clusters  of 1d defects, or trajectories.
Another Abelian example is the $Z_2$ gauge theory.
In this case the confining configuration is a percolating cluster of surfaces, or vortices.

In the non-Abelian
case \footnote{For simplicity we consider  pure SU(2)
gauge theory, without matter.} there is no consensus yet
on the particular choice of the magnetic degrees of freedom.
The general idea is to reduce the non-Abelian degrees of freedom
to the pattern of Abelian theories where the mechanism of confinement is
 well understood.
 For this purpose one introduces the
so called projected fields:
\begin{equation}\label{projected}
A_{\mu}^a(x)~\to~A_{\mu}^3(x)~, or~~A_{\mu}^a(x)~\to~(Z_2)_{\mu}(x)~~,
\end{equation}
where $A_{\mu}^3(x)$ is an Abelian gauge field, and $(Z_2)_{\mu}(x)$
is a $Z_2$ gauge field ($(Z_2)_{\mu}(x)=\pm 1$ for links).
The U(1) projections emphasize the role of the monopole trajectories
or 1d defects while $Z_2$ projections allow to find center vortices,
or 2d defects \footnote{The literature on the confinement is vast.
For the purpose of orientation and further references we can mention
reviews in Ref. \cite{reviews} and in Ref. \cite{greensite} exposing in detail
the $U(1)$ and $Z_2$ projections, respectively.}.

The projected fields (\ref{projected}) are related to the original
Yang-Mills fields in a nonlocal way and
theoretical interpretation of the lattice data is obscured by their  use.
It turns impossible to reconstruct configurations
of the original non-Abelian fields $A_{\mu}^a(x)$
which correspond to the monopoles or vortices.
 However, it seems indisputable that effectively
Abelian degrees of freedom
 are responsible for the confinement.

In this note we will reverse the problem and instead of trying to directly
interpret the lattice data  consider  classification of
non-Abelian singular fields in the continuum which could be
responsible for violations of the Bianchi identities while not
inducing violations of the color conservation. We argue that a
natural candidate for such defects are singular fields living on
two-dimensional surfaces. The central point is that such fields are
Abelian in nature and, therefore, violations of the Bianchi
identities and of the equations of motion are disentangled.

As far as the  mathematical tools are concerned,
we utilize
the so called surface
operators  \cite{gukov} which describe four-dimensional (4d)
non-Abelian fields living on 2d surfaces.
It is interesting that from the phenomenological
point of view the use of the surface operators suggests a unification
of the two alternative Abelian pictures mentioned above. Namely,
the defects are 2d, as emphasized by the center projections, while
the fields living on them are Abelian, as
commonly emphasized by the $U(1)$ projections.

\section{Abelian case}

 To set up the framework, let us review first
the compact U(1) theory \cite{polyakov}. The Lagrangian is the
same as for a free electromagnetic field:
\begin{equation}\label{original}
L_{U(1)}~=~{1\over 4 e^2}(F_{\mu\nu})^2~~,
\end{equation}
supplemented, however, by the condition that the Dirac string carries no action.
The condition is automatically satisfied in the lattice version
of the theory.

Admitting singular fields, or monopoles into the theory violates Bianchi identities
and the modified Maxwell equations now read:
\begin{equation}\label{current}
\partial_{\mu}F_{\mu\nu}~=~0~~, ~~\partial_{\mu}\tilde{F}_{\mu\nu}~\equiv~j_{\nu}^{mon}~,~~
\partial_{\nu}j_{\nu}^{mon}~=~0~.
\end{equation}
  where $j_{\nu}^{mon}$ is the monopole current.

The non-vanishing,
 conserved current $j_{\nu}^{mon}$ can be traded for a magnetically
charged scalar  field $\phi_M$. This is a
generic field theoretic phenomenon (in the Euclidean space-time).
The derivation can be found in quantum-geometry courses,
see, e.g.,  \cite{ambjorn}
while specific applications to lattice monopoles are discussed,
in particular, in Refs \cite{stone}.
Here we will remind, for a later use, a few basic steps
in  relating the monopole current to a magnetically charged field.

Observing $j_{\mu}^{mon}$, say, in lattice simulations is equivalent
to observing particle trajectories. Therefore, it is reasonable to
start from the so called polymer formulation of a free scalar field
theory with the action
\begin{equation}\label{polymer}
S_{cl}~=~M_{bare}\cdot L~~,
\end{equation}
where $M_{bare}$ is the (bare) mass and
$L$ is the length of trajectory.

The mapping of the polymer representation (\ref{polymer}) into the standard
field theoretic representation is achieved through evaluating the path integral
for the particle propagator:
\begin{equation}\label{sum}
D(x,y,M)~=~\Sigma_{paths}\exp\big(~-S_{cl}(x,y,M)\big)~~.
\end{equation}
The sum (\ref{sum}) can be evaluated exactly and one establishes \cite{ambjorn}
a relation between the physical mass $m^2_{\phi}$ and the polymer-representation
mass $M$:
\begin{equation}\label{finetuning}
m^2_{\phi}~\approx~{const\over a}\big(M_{bare}(a)~-~{{\ln 7}\over a}\big)~~,
\end{equation}
where $\ln 7$ is a geometric factor specific for a cubic lattice in $d=4$
and we introduced explicitly dependence of the bare mass (\ref{polymer})
on the lattice spacing $a$ which is an ultraviolet cut off.

To relate the polymer approach to the physics of the lattice monopoles
one identifies the bare mass $M_{bare}$ in (\ref{polymer}) with
  the radiative mass of the monopole:
\begin{equation}
M_{bare}~\to~M_{mon}~=~{const\over a e^2}~~,
\end{equation}
where $e^2$ is the electric charge
squared.
The Higgs, or confining phase corresponds to $m^2_{\phi}<0$. Once $m^2_{\phi}=0$
there emerges an infinite, or percolating cluster of the monopole
trajectories. The relation between the monopole density $\rho_{mon}$
and the field-theoretic
vacuum expectation value reads as:
\begin{equation}\label{vev}
\langle 0||\phi|^2|0\rangle~=~const\cdot a\rho_{mon} ~~,~~ (\langle
0|\phi|0\rangle)^2~=~const\cdot a\rho_{mon}^{perc}~~,
\end{equation}
where $\rho_{mon}$ is the total monopole density and
$\rho_{mon}^{perc}$ is the density of the percolating monopoles.

\section{Non-Abelian singular fields}

The Abelian construction just described does not generalize to
the non-Abelian case. Indeed, if
$$(D_{\mu}\tilde{G}_{\mu\nu})^a~=~j_{\nu}^a~~,$$
where $j_{\nu}^a$ is the monopole  current, then
the colored current $j_{\nu}^a$
would modify also the equations of motion, not only the Bianchi identities
since any colored current is a source of gluons.
Moreover, if we trade the current $j_{\nu}^a\neq 0$ for a scalar field
then this field is colored, $\phi^a$ and its vacuum expectation value would violate color conservation.

Now, we are coming to a crucial point. We do not take these
difficulties as a proof that singular
fields have no role to play in the non-Abelian
case. Instead, we merely conclude that trajectories, or
1d defects are not adequate to the non-Abelian case and
will be looking for defects of other dimensions. Note that the fact
that monopoles are intrinsically U(1) (not $SU(2)$)
 objects has been
emphasized since long, see, in particular, \cite{coleman}.

Turn now to two-dimensional defects, or surfaces. Classification of
singular field living on a surface, or 2d defects
is actually contained in \cite{gukov} \footnote{The
observation on Abelian nature of the non-Abelian fields living on a
surface, crucial for our considerations, was intensely exploited
also earlier \cite{ch}. However, the surface operators also
allow to ascribe to the surfaces density of topological charge,
see below,
and phenomenological consequences from this observation
have not been  considered, to our knowledge.}. The central point is that non-Abelian
fields living on a surface can in fact
  be rotated to an Abelian direction and as a result  violations of the Bianchi identities
($D_{\mu}\tilde{G}_{\mu\nu}\neq 0$)  can be consistent
 with the validity of the equations of motion
($D_{\mu}G_{\mu\nu}=0$).

Introduce first the action associated with the surfaces in the form:
$$S_{surface}~=const \int d\sigma_{\mu\nu}G_{\mu\nu}^a~~ ~.$$
This action is not gauge invariant. However, it can be redefined
 in such a way as to respect the
non-Abelian invariance.  The reason is that the surface interaction
 at each point $x$ involves only a single component of
the field strength tensor $G_{\mu\nu}^a$. Therefore, one can use
gauge invariance to rotate this particular component   to the Cartan
subgroup:
\begin{equation}\label{rotation1}
d\sigma_{\mu\nu}G_{\mu\nu}^a(x)~\to~
d\sigma_{\mu\nu}G_{\mu\nu}^3(x)~,~~
\end{equation} where for simplicity we considered the gauge group
$SU(2)$.

Note that the projection (\ref{rotation1}) is determined up to a
sign. One can fix the sign by imposing the condition
\begin{equation}\label{gauge}
\big(d\sigma_{\mu\nu}G_{\mu\nu}^3\big)(x)\cdot
\big(d\sigma_{\mu\nu}G_{\mu\nu}^3\big)(y)~>~0~~.
\end{equation}
As is argued in \cite{gukov} the surface can be endowed also with a
dual field $\tilde{G}_{\mu\nu}^a$ which can also be rotated to the
Cartan subgroup:
\begin{equation}\label{rotation2}
d\sigma_{\mu\nu}\tilde{G}_{\mu\nu}^a(x)~\to~
d\sigma_{\mu\nu}\tilde{G}_{\mu\nu}^3(x)~,~~ \end{equation} where the
point $x$ belongs to the  surface. Note that only
one of the two rotations (\ref{rotation1}), (\ref{rotation2}) is
uncertain in sign.

From the point of view of the lattice formulation, the possibility
of two independent (apart from a sign) rotations assumes
a particular regularization. Indeed, the dual field is defined on
the dual-lattice plaquettes and in this sense can be rotated
independently. However, such a procedure would assume simultaneous
use of both direct and dual lattices which is not necessarily
legitimate. We will not go into details of this issue here and just
follow Ref. \cite{gukov} in
postulating,  in the continuum-theory language, existence of  closed surfaces, with surface element $d\sigma_{\mu\nu}$,
and with the fields $G_{\mu\nu}^3,\tilde{G}_{\mu\nu}^3$
associated with the surfaces.
One ascribes then to the surface the following action
\begin{equation}\label{actionn}
S_{surface}~=~\alpha\int d\sigma_{\mu\nu}G_{\mu\nu}^3~+~\beta\int d\sigma_{\mu\nu} \tilde{G}_{\mu\nu}^3~~,
\end{equation}
where $\alpha,\beta$ are constants.

Note that in terms of invariants the two-dimensional defects
considered have both non-vanishing action and
topological-charge densities,
\begin{equation}\label{itog}
G^2(x)~>~0,~~G\tilde{G}(x)~\neq~0~~,
\end{equation}
as far as the point $x$ belongs to the  surface.
The fields $G_{\mu\nu}^3, \tilde{G}_{\mu\nu}^3$ can be directly
defined in terms of the invariants as
\begin{equation}\label{definition}
G^3_{\mu\nu}(x)~\equiv~+\sqrt{G^2(x)}{d\sigma_{\mu\nu}\over |d\sigma_{\mu\nu}|}~,~~
\tilde{G}^3_{\mu\nu}(x)~\equiv~{G\tilde{G}(x)\over \sqrt{G^2(x)}}{d\sigma_{\mu\nu}\over |d\sigma_{\mu\nu}|}~,
\end{equation}
where the point $x$ belongs to the surface.

\section{Various facets of defects}
\subsection{Wilson lines}

Two-dimensional surfaces introduced above can be considered on classical
or quantum level, as external objects or as dynamical degrees of freedom.
Indeed, turn first to the example of  the Wilson   line.
Classically (and in Abelian case)
$$ \int_{C}A_{\mu}dx_{\mu}~=~\int_{surface}d\sigma_{\mu\nu}H_{\mu\nu}$$
where the surface is spanned on the contour $C$. In the classical case
 the contour integral  merely 'measures' the external magnetic flux.
If the potential $A_{\mu}$ is given locally by a pure gauge, the
contour integral is quantized.

If we introduce the Wilson line as an external object and consider
it quantum mechanically, then there emerge divergences due to the self
energy:
\begin{equation}\label{uvwilson}
\langle Tr P\exp\{-\int_{C}\hat{A}_{\mu}dx_{\mu}\}\rangle~\sim~
\exp(-const~g^2 L/a)~,
\end{equation}
where $L$ is the perimeter of the Wilson line $C$, $a$ is the lattice spacing,
$g^2$ is a coupling and we keep only the most divergent
piece.

Finally, one can consider the loops as dynamical objects, generated within
the theory itself. In particular, it is quite common nowadays to consider a condensate
  of Polyakov's lines \cite{polyakov1} which are defined at finite temperature:
$$ Tr~L~\equiv~Tr P\exp\{-\int_0^{1/T} dx_0\hat{A}_0({\bf x})dx_0\}~~$$
and are the Wilson lines stretched in the time direction.

Phenomenologically, it is appealing to assume that there exists effective potential for
the vacuum expectation value of the Polyakov's lines:
\begin{equation}
V(<tr L>)~=~-c_2<tr L>^2~+~c_4<Tr L>^4~~,
\end{equation}
where in the deconfining phase, or at $T>T_c$ the constants $c_{2,4}$ are positive, see, in particular \cite{pisarski}.

Let us emphasize that for the vacuum expectation value $<Tr L>$ to be non-vanishing,
$$<Tr L>~\neq~0~~,$$
  the   ultraviolet
divergence exhibited by  (\ref{uvwilson}) is to be canceled by the entropy.
We discussed above such a cancelation on the example of theory of a free particle
within the polymer approach.

\subsection{Surface integrals classically}

Surface integrals similar to  (\ref{actionn}) appear in the GNO classification \cite{goddard}
of the non-Abelian monopoles \footnote{The surfaces are closed in this case.
Hereafter we will consider closed surfaces in all the cases.} . One assumes that
there exist classical
 solutions such that at large distances $r$ from the position of the
monopole the  (space-space) components  of the gauge field tensor
take the form
\begin{equation}\label{magnetic}
G_{ij}~=~{\epsilon_{ijk}r_k\over r^3}G(r)~~,
\end{equation}
where $G(r)$ are magnetic charges. The solution can possess an electric charge as well and then the dual field $\tilde{G}_{ij}$
is also non-vanishing.
The observation \cite{goddard} is that in gauge theories, though electric charge takes values in the weight
lattice of the gauge group, the (quantized) magnetic charge $G$
 takes values in the weight lattice
of a dual group.

In case of the gauge group $SU(N)$ the dual group
is $SU(N)/Z_N$ where $Z_N$ is the center group. In the simplest case of SU(2) group which
we concentrate on  the magnetic charges $Q_M=\pm 1$
are to be identified
since the corresponding magnetic fields (\ref{magnetic}) can actually be gauged transformed into each other.

\subsection{Surface operators as external objects}

As is mentioned above,
Ref. \cite{gukov} introduces the surface operators as external objects
on the quantum level.
One
can measure  vacuum expectations value of the surface operators by substituting the
the vacuum fields (\ref{definition}) defined for each configuration on the lattice and then averaging
over the configurations.

Clearly, the vacuum expectation value is suppressed by the ultraviolet divergent
self energy
proportional to the area of the closed surface:
\begin{equation}\label{uvsurface}
\langle S\rangle~\sim~\exp\big(-\alpha\cdot (const) (Area)/a^2\big)~~,
\end{equation}
where $\alpha$ enters definition of the action (\ref{actionn}) and
$(const)$ is related to the average value of the plaquette
action.
In other words, $G^3_{\mu\nu}(x)$ living on the surface appears
singular in the limit of the vanishing lattice spacing $a\to 0$.
The divergence corresponds to self energy of colored dipoles
living on the surface and oriented in the third direction
of the color space. It is an analog of the divergence
(\ref{uvwilson}) in case of the Wilson lines.

One can measure also density of the topological charge
$G\tilde{G}(x)$ for the points $x$ belonging to the surface,
~$x\epsilon \Sigma$. The results of the measurements
would depend on the properties of the Yang-Mills vacuum
and cannot immediately be predicted.

As in case of the Wilson line one can consider
quantities not sensitive to the ultraviolet
divergence (\ref{uvsurface}).
For this purpose one can consider, for example, closed
surfaces  unifying two planes separated by distances $r$
and measure potential energy as function of $r$.
Studying expectation values of the surface operators
could probe the confinement, similar to the case
of the Wilson lines \cite{gukov}.
Indeed, the vacuum expectation value
of the surface operator
(apart from the just mentioned ultraviolet divergent piece)
 might exhibit either volume or area laws:
\begin{equation}\label{volume}
\langle S\rangle
\sim \exp\big(-(const)(Volume)\big)~,~or~
 \langle S\rangle \sim\exp\big(-(const)(Area)\big)~.
 \end{equation}
 Note that
by measuring the vacuum expectation values of the surface operators
one studies the interaction of particles having color magnetic and
dipole moments, but no color charge. Therefore, it is our guess
that there is no volume law even in the confining phase.

\subsection{Surfaces as dynamical defects}

In this paper we are going to make the next step and assume that
surfaces become dynamical and condense in the vacuum in the
confining phase so that  there is a net of surfaces endowed with   fields living
on them. The picture is similar to condensation of the
Polyakov's lines in
the deconfinement phase.

It is a strong assumption concerning the dynamics of non-Abelian theories
which in no way can be derived from first principles.
Our assumption is motivated by the lattice data
that do demonstrate existence of 2d surfaces
percolating through the vacuum
\cite{greensite,zakharov}. For a more theoretically oriented reader
let us mention that holographic models predict
both percolation of the Polyakov's lines in the deconfining phase
and percolation of magnetic  strings in the confining phase
\cite{gorsky}.

In more detail, we will rely on the Sakai-Sugimoto
holographic model \cite{sakai} which is most successful in describing low-energy QCD.
The metrics of this model is formulated in terms of the ordinary 4d space
$(t, x^i, i=1,2,3)$, extra fifth dimension $u$ with horizon at $u=u_{\Lambda}$,
one more extra compact dimension $x^4$ and compact sphere $\Omega_4$:
\begin{eqnarray}\label{metrics}\nonumber
ds^2=\big({u\over R_0}\big)^{3/2}(-dt^2+\delta_{ij}dx^idx^j+f(u)dx_4^2)+
\big({u\over R_0}\big)^{-3/2}\big({du^2\over f(u)}+u^2d\Omega_4^2\big)\\
f(u)~=~1~-~\big({u_{\Lambda}\over u}\big)^3.
\end{eqnarray}
It is also crucial that the defects wrapped on the compact dimension $x^4$
have non-trivial $\theta$-angle dependence.
Note that the metric in dimensions $(x^4,u)$ has a cigar form so that
the radius of the $x^4$ compact dimension tends to zero at the horizon $u=u_{\Lambda}$.
Which means that the defects wrapped around this dimension have tension
vanishing at the horizon. In particular, the magnetic strings are defined
as D2 branes wrapped around the $x^4$ dimension. They look as strings, or 2d surfaces in
the ordinary 4d space and, in case of large enough area, have vanishing (classically) tension.
Because the D2 branes are wrapped on the compact $x^4$ dimension
one expects that they carry density of the topological charge.

The metrics (\ref{metrics}) refers to the case of zero temperature $T=0$
which we are mostly interested in. At finite T it is convenient to
use Euclidean space, with a compact time direction
$\tau$. Thus, at $T\neq 0$ there are two
compact coordinates, $x^4$ and $\tau$. The deconfining phase transition
is then understood as change of geometries in two-dimensional $(x^4,u)$ and $(\tau,u)$
subspaces. Namely, at $T>T_c$ the cigar-shape geometry sets in the $\tau,u$
coordinates while in the $(x^4,u)$ coordinates the geometry becomes cylinder-like.
As a result, defects wrapped in (Euclidean) time direction
become tensionless and could condense.
As is argued in \cite{gorsky} this is a general phenomenon within the
dual model: below $T_c$ defects condensed in the vacuum have nontrivial
$\theta$-angle dependencies while at $T>T_c$ it is the defects wrapped on the
compact time direction which are tensionless. The Polyakov line is wrapped
around the time direction and  becomes tensionless at $T>T_c$
according to the dual model. This fits well the phenomenological models
which postulate condensate of the Polyakov's lines.
The magnetic D2 branes are wrapped around the compact $x^4$
direction and are tensionless at small temperatures.
They are expected to condense at small T.

\section{Surface operators and magnetic branes}

The hint which we get from the dual models is that in the Yang-Mills vacuum
there could exist magnetic strings, tensionless. We conclude that their tensionlessness  is
to be 
manifested as percolation of  2d surfaces, or existence of an infinite
cluster of such vortices. As usual in quantum geometry, the surfaces are to be endowed
with action. Moreover, the  vanishing string tension implies action-entropy balance.
Another important prediction of the dual models is that the surfaces are
carrying topological charge density (since the D2 branes are wrapped 
around the compact $x^4$ coordinate).

A crucial observation is that the surface operators are an adequate
mathematical tool to consider the magnetic strings because they also
describe 2d surfaces endowed both with action and topological charge.
In this section we will consider further phenomenological consequences
of such identification.

\subsection{Non-Abelian 'monopoles'}

Assumption on the dynamical nature of the 2d defects brings about
a conclusion that there should exist also 1d defects living
on surfaces.
Indeed, consider first two external surfaces with
$$(G^2)_1~=~(G^2)_2 ~,~(\tilde{G}G)_1~=~-(\tilde{G}G)_2~.$$
As far as the surfaces are treated as external objects
they are different although degenerate in energy. However,
once we allow for dynamical surfaces to exist
the two degenerate states mix up
 and the
actual ground state would consist of  their mixture.
Lines on which the $\tilde{G}G$ changes its sign are one-dimensional
defects. Which we will call non-Abelian monopoles (since they are associated
with lines, or trajectories).

There is an important difference in the structure of gauge fields associated with
the Abelian and non-Abelian monopoles.
Consider the Abelian case first and
introduce a  time slice of the 4d space.  Then at some
point a singular magnetic field may not have a particular direction:
\begin{equation}\label{sphere}
{\bf H}_{Abelian}~\sim~{{\bf r}\over r^3}~~,
\end{equation}
where ${\bf r}=0$ is the position of the singularity. Such a
singularity corresponds to the Dirac monopole.

In the non-Abelian case consider a time slice of the surface which is a line,
with coordinate $\tau$. At some point $\tau_0$ the magnetic field
living on the line may change direction:
\begin{equation}\label{line}
{\bf H}_{non-Abelian}~=~ |{\bf
H}|{(\tau-\tau_0)\over|\tau-\tau_0|}~~.
\end{equation}
It is crucial that the magnetic field in the non-Abelian case is not
spherically symmetric but is line-like. Such field configuration is
allowed by the non-Abelian invariance as a defect while the
spherically symmetric field (\ref{sphere}), familiar from the
Abelian case, is not allowed.
If we now include development in time both the Abelian and
non-Abelian monopoles become lines, or trajectories. However, in the
Abelian case the trajectories live on 4d space while in the
non-Ableian case they live on   2d surfaces.  

Let us emphasize that the very existence of the 1d defects depends crucially
on the assumption that the 2d defects (surfaces) carry both fields
$G^3_{\mu\nu}$ and $\tilde{G}^3_{\mu\nu}$.
If there were only magnetic fields living on the surface
the change of the direction of the magnetic field, see Eq. (\ref{line})
would be obscured by the non-Abelian gauge invariance which allows to change the sign
of the magnetic field.  

\subsection{Open strings}

So far we were discussing closed surfaces, or strings. One might ask
whether it is possible to have an open string as a defect. Thus, we
wish to introduce a 1d boundary. For the whole construction to be
non-Abelian invariant, this boundary, or line is to have also
non-Abelian invariant meaning. Physicswise, the magnetic field which
exists along the strings 'flows' to its end in case of an open
string. Thus, it is natural to have a monopole at the end points.
Moreover, this 'monopole' is to be defined in an $SU(2)$ invariant
way. There is only a single monopole of this type, that is the $Z_2$
monopole \cite{lubkin}, or the 't Hooft line \cite{thooft1} (in case
of $SU(N)$ group there are $(N-1)$ species of  $Z_N$ monopoles).

The $Z_2$ monopole can be viewed also as an ordinary Abelian
monopole with magnetic flux $2\pi/g$:
\begin{equation}\label{flux}
\int_{Z_2~monopole} d\sigma_{\alpha\beta}G_{\alpha\beta}~=~{2\pi\over g}
\end{equation}
where $g$ is the gauge coupling. Note that the corresponding Dirac
string is having quadratically divergent action and in this sense is
visible (for further discussion see  \cite{gubarev}). Finally, we
note that the magnetic field living on the 2d surface is not
quantized and the magnetic flux flowing to the end points of the
string  is generically not equal to the flux (\ref{flux}).

\section{Lattice data}

As we have already mentioned, the lattice data
demonstrate existence of 2d surfaces with properties
close to the predicted above. Mini-reviews of the
data can be found, e.g., in Ref. \cite{zakharov}.
For the sake of completeness, we reiterate main points here.

Definition of the surfaces on the lattice are algorithmic
and is difficult to translate into the continuum-theory
language. This is the main difficulty in interpretation of the
data. As a partial compensation for this difficulty the lattice
itself allows for specific and powerful tests of the hypothesis
that some particular lattice defects have meaning in the continuum
limit. First of all, one can check scaling laws
which are obeyed by physical objects,
for reviews see \cite{greensite,reviews}.

Let us give an example.
Imagine that there is a technical definition of closed
2d surfaces in Yang-Mills vacuum. One can measure global characteristics of it.
In particular, one measures the total area of the surface.
Imagine, furthermore, that one finds:
\begin{equation}\label{area}
(Area)_{tot}~\approx~\Lambda^2_{QCD}V_{tot}~~,
\end{equation}
 where $V_{tot}$ is the total volume of the lattice.
 The proportionality to $V_{tot}$ is triviality for large
 enough volumes. However, the proportionality to the
 physical scale, $\Lambda_{QCD}^2$ is highly nontrivial.

 On the lattice, one changes the lattice spacing $a$ and
 the bare coupling $g^2(a)$, in accordance with the renormgroup.
Absence of explicit dependence of the area (\ref{area})
 on the lattice spacing implies strong 'conspiracy' of the
 data. Indeed
 the statement (\ref{area}) can be rewritten
 as observation that probability of a given plaquette to
 belong to the surface is proportional to
 \begin{equation}\label{probability}
 \theta_{plaq}~\approx~const (a\cdot\Lambda_{QCD})^2~~,
 \end{equation}
 and changing the lattice spacing by, say, factor of 2
 changes the probability (\ref{probability}) by a factor of 4
 which is a well defined prediction and a large effect
 numerically.

 Once relation (\ref{area}) is found empirically,
 in terms of a certain $Z_2$ projection of the original
 non-Abelian fields   it becomes a strong evidence
 that using this projection allows to detect a gauge
 invariant physical object. Indeed, the property (\ref{area}) is perfectly
 gauge invariant. Moreover, it implies that the tension of the
 strings is in physical units:
 \begin{equation}\label{tension}
 T_{string}~\sim~\Lambda_{QCD}^2~~,
 \end{equation}
 since it is the tension which controls the area.
 Of course it is not  a complete proof yet
 of the physical nature of the surfaces. Moreover, one could reverse the
 procedure
and just try to define such a projection that Eq. (\ref{area})
holds with good accuracy. Thus, we trade the property
(\ref{area}) for the definition of the surface.
The central problem then,  are there further gauge invariant
properties possessed by the surface. In case of the
center vortices the answer is in positive (for review and
details see \cite{greensite,zakharov}).

In particular, there is an extra non-Abelain action associated with the
surfaces \cite{gubarev}:
\begin{equation}\label{a2}
\delta S_{surface}~\approx~(const) {(Area)\over a^2}~~,
\end{equation}
where the use of the symbol `approximate' always means that
the evidence is numerical, i.e., true within some error bars.
We do not discuss the error bars here and only sketch the overall
picture.

The physical meaning of (\ref{a2}) is that the construction is consistent
at short distances. Indeed, in the language of the quantum geometry
having (\ref{tension}) is consistent only if there is action-entropy
balance:
\begin{equation}
(Tension)(Area)~=~(Action)_{surface}~-~(Entropy)_{surface}
\end{equation}
where both the action and entropy associated with the surface
are ultraviolet divergent while their difference is finite.

Another confirmation of this amusing picture is provided by measurement
of the extra action associated with plaquettes next to those belonging
to our surfaces. It vanishes identically:
\begin{equation}
\delta S_{next~to~surface}~=~0~~.
\end{equation}
In physical terms, it is a confirmation that the action
is indeed  associated with the field living on the surface
as we have been conjecturing.

A crucial property of the surfaces we are discussing is that
they are expected to possess not only action but density of
the topological charge as well. Amusingly enough, this prediction
has already been tested and, within error bars, confirmed.
Namely, one measures on the lattice for each particular configuration
space distribution of wave functions of topological modes of
quarks. It turns out that the intensity of the topological charge
(as measured by intensity of the topological fermionic modes)
is correlated positively with the position of the surfaces
and the correlation grows as an inverse power of the lattice spacing
\cite{kovalenko}:
\begin{equation}\label{correlator}
 \{(Intensity~(topological~ modes), ~~Intensity~(surfaces)\}~\sim~a^{-\gamma}~~,
\end{equation}
where the critical exponent $\gamma\approx 2$ \footnote{The precise
definition of the correlator (\ref{correlator}) is rather technical
in nature. It can be found in the original papers \cite{kovalenko}.}.

Apart from the surfaces, or 2d defects one observes on the lattice
1d defects, or monopole trajectories as well.
The monopoles
observed on the lattice \footnote{So called monopoles of the
Maximal Abelian Projection.} do seem to have properties close
to those predicted theoretically. Namely, they lie on the surfaces just
discussed and their non-Abelian field is rather line-like,
aligned with the surfaces.

 \subsection{Localization of classical solutions}

Thus, the center vortices on the lattice do have properties similar
to those inherent to the surfaces discussed in the preceding sections
and based on the surface operators. However, the reader might
feel uncomfortable to discuss confinement in terms of singular fields
(\ref{uvsurface}).
Indeed, in the continuum theory one usually discusses non-perturbative physics in Yang-Mills theory in terms of soft fields. On the other hand, the lattice
data reveals that hard fields are associated
  with the confinement and this cannot be disregarded. Thus, we have a
paradox.

It is worth emphasizing therefore that it
would be wrong to consider the infrared and ultraviolet sensitive
descriptions as mutually excluding each other.
It is much more adequate to think in terms
of a kind of duality, that the nonperturbative physics can be described
either in infrared or ultraviolet terms. Moreover, one description
probably smoothly goes into the other one as a functions
of resolution of measurements, or of the value of the lattice
spacing \cite{jubileum}. The term 'resolution' is used here because
generating each filed configuration on the lattice,
$\{A_{\mu}^a(x)\}$ can be considered as a measurement of
the gauge fields on the whole lattice with resolution of the lattice spacing $a$.

Unfortunately,
there is no detailed understanding of the role of the resolution.
Probably, one deals with quite a general phenomenon which
can be called stabilization of classical solutions
with high-frequency oscillating fields. At this time the guess
can be substantiated, however, only by some analogies
and intuitive considerations.
It would be extremely interesting to clarify further these issues
  \footnote{Analytically, interplay between the infrared and ultraviolet sensitive
 descriptions was discovered first in 2d CP(N) models \cite{fateev}.
 Namely, it was proven that upon summation over all the instanton solutions
 (including multi-instantons) the instanton ensemble is described
 by the Coulomb-gas model. The constituents are point-like and possess
 topological charge equal to $1/N_c$ fraction of that of the instanton.
 }.

Let us start with a far fetched analogy from
classical mechanics. Consider a pendulum with
  frequency
$\Omega_{slow}$. Apply now a fast oscillating force
to
the fixed point of the pendulum, with amplitude
$a_{fast}$ and frequency $\omega_{fast}$. Then the
effective potential which determines the resulting motion of
the pendulum becomes:
\begin{equation} \label{effect}
U_{eff}~=~mgl(1-\cos\phi)+a^2_{fast}\omega^2_{fast}\sin^2\phi
\end{equation}
where $\phi$ is the angle of deviation of the pendulum
from the vertical line.

If the ratio
$\omega_{fast}/\Omega_{slow}$ is large
the effective potential (\ref{effect})
is much steeper than the original one.
The resulting amplitude of oscillation of
the pendulum around $\phi=0$ is much smaller
than without the oscillating force.
Moreover, the point $\phi=\pi$ becomes a minimum
of the effective potential as well \footnote{This is a famous
example due to S. Kapitza, see L.D. Landau and E.M. Lifshitz,
Course of Theoretical Physics, vol. 1.}.

In the Yang-Mills case, the analogy to slow motion
is provided by a classical solution, say instanton, of
size $\rho_{inst}\sim\Lambda_{QCD}^{-1}$.
Imagine that we would like to next include effect
of zero-point fluctuations
which are fast oscillating fields. The zero-point fluctuations
are dominated by the ultraviolet scale:
$$[a_{\mu}^a]_{quant}~\sim~a^{-1}_{latt}$$
where $a_{\mu}^a$ is the quantum gauge field and $a_{latt}$
is the lattice spacing.

The classical instanton solution $(A^a_{\mu})_{class}$
is no longer a solution. Indeed $(A^a_{\mu})_{class}\sim \rho_{inst}/g$
and is much smaller, by the factor
of order $(a_{latt}/\rho_{inst}) $ than the quantum field $a_{\mu}^a$.
In analogy with the mechanical example mentioned above
one should rather first average over fast quantum oscillations
and derive an effective potential.

Of course, there are no means to solve this problem analytically.
However, it seems rather obvious that the 'instanton' shrinks,
the same as the motion of the pendulum.
Indeed, instantons are in fact  generic non-perturbative fluctuations
and the probability to find such a fluctuation is of order
\begin{equation}\label{nonpert}
\theta_{non-pert}~\sim~\exp\big(~-~(const)/\alpha_s(a^2_{latt})\big)~~,
\end{equation}
where $\alpha_s(a^2_{latt})$ is the running coupling.
The crucial point is that $\alpha_s$ is normalized on the scale
on which measurements are performed, i.e. on the scale $a_{latt}$.

 Estimate (\ref{nonpert}) implies that the instanton volume shrinks to zero
 as a power of the lattice spacing $a_{latt}$. Moreover,
 relying bluntly on the analogy with the pendulum,
 we come to a bizarre conclusion that
\begin{equation}\label{naive}
V_{instanton}~\sim~(const)\cdot a^4_{lattice}~~.
\end{equation}
Strange to say, but this guess might well be true, according to the lattice data
\footnote{
The phenomenon of the shrinking of classical solutions was discovered by the  ITEP
lattice group \cite{gubarev}. However, the classical solutions relevant to this paper belong
to a dual formulation of the Yang-Mills theories and the status of these
solutions is rather vague. The idea to check this phenomenon on the case
of 'instantons', or chiral defects belongs to the
authors of the first paper in Ref. \cite{deforcrand}.}.

Let us specify the notion of the 'volume occupied by an instanton'.
The definition is provided actually by considering zero- or near-zero fermionic modes.
In the classical instanton case the topological modes are located  in the same region of
the 4d space as the instanton itself:
\begin{equation}\label{ferm}
 |q_{\lambda=0}(x)|^2~\sim~{\rho^4\over (\rho^2+x^2)^3}~~,
\end{equation}
where $q_{\lambda=0}$ is a zero mode of the Dirac equation
for a quark in external gluon field:
$$\hat{D_{\mu}}\gamma_{\mu}q_{\lambda=0}~=~0~,$$
and to derive (\ref{ferm}) one substitutes the instanton field for
the external gluon field.

On the lattice, one can solve the Dirac equation numerically
substituting into the covariant derivative $D_{\mu}$
the gluon fields generated on the lattice for a particular
configuration $\{A_{\mu}^a(x)\}$.
 Then the volume occupied by the
 topological fermionic modes can be understood
 as the volume occupied by `instanton modified by high-frequency
 perturbative fields'. Lattice data rather support estimate (\ref{naive}).

 Let us emphasize that the shrinking volume of the instanton
 implies also singular fields inside the small region occupied
 by the modified instanton.
 Indeed, the average value of the topological charge squared,
 or topological susceptibility,
 \begin{equation}
 \int d^4x <0|~G\tilde{G}(x),~G\tilde{G}~|0>~\sim~\Lambda_{QCD}^4~~,
 \end{equation}
 represents a matrix element and its value cannot depend on details of the
 measurements, in particular on the value of $a_{latt}$. On the other hand,
 the value of $Q_{top}^2$ for each configuration is equal to the number
 of exact zero modes of the quarks. Thus, the total topological charge
 of configurations cannot depend on the shrinking of each particular
 lump of the topological charge.
 Hence, the topological fields become singular in the limit
 of the vanishing volume of the lumps.

 Assuming all this to be correct, what are implications for
 interpretation of 2d surfaces? To answer the question
 note that in terms of the dual model instantons are D0 branes.
 These branes have one extended coordinate which is wrapped around
 the $x^4$ circle. In the geometrical language of the dual models
 (\ref{metrics}) this wrapping is responsible for the non-trivial
 topological charge of instantons. To summarize: the instantons
 are points in ordinary 4d space, within the defects classification
 of the dual models.

 On the other hand, in the ordinary field theoretic language
 the instantons are rather extended objects of the size of order
 $\rho\sim\Lambda_{QCD}^{-1}$. Remarkably enough, the shrinking
 caused by $a_{latt}\to 0$ just brings an instanton to a point:
 \begin{equation}
 \Big(G^2\sim {\rho^4\over (\rho^2+x^2)^4}\Big)~\to ~\Big(G^2\sim \delta^4(x)\Big)~~,a_{latt}\to 0.
 \end{equation}
 Thus, one can speculate that in the limit of vanishing lattice
 spacing the geometry of the classical solutions becomes
 similar to the geometry of the classical solutions
 in the dual formulation.

 Coming back to our surfaces populated with singular non-Abelian
 fields they could well be a high-frequency image of classical
 solution of Yang-Mills theory in the dual representation.
 This guess cannot be supported however
 by any straightforward derivation.

\section{Open questions. Conclusions}

As we have demonstrated above, different approaches to magnetic
degrees of freedom converge to a common and highly non-trivial
picture. In particular, the dual models \cite{gorsky}
relate percolation of the magnetic strings to the prediction
that
 they should carry density of the topological charge.
The surface operators \cite{gukov} provide us with an adequate
construct to describe such surfaces endowed with both gauge fields and their
duals. On the lattice, one does observe percolating surfaces, with tension
in physical units and endowed with topological charge.

In this note we emphasized that the monopole mechanism of
confinement, whatever it means in detail, assumes   significant role
to be played by singular non-Abelian fields. Indeed, 'electric'
degrees of freedom are introduced directly in the Lagrangian, while
'magnetic degrees of freedom' are emerging through violations of
Bianchi identities. Hence magnetic degrees of freedom are associated
with singular fields.

The construction is well understood and known since long
\cite{polyakov} in the Abelian case. One can either  start with a
gauge field  plus a  scalar, Higgs field, or one can start with pure
gauge field but admit singular field configurations, monopoles. Then
these singular fields, or defects can be traded for a scalar field
and one comes  back to an effective Higgs-like formulation.

In the non-Abelian case monopoles, or 1d defects do not match the
group structure  \cite{coleman,lubkin}. Instead, one should look for
classification of singular fields consistent with the non-Abelian
symmetry. As is argued in fact recently \cite{gukov} such defects
are 2d dimensional surfaces with  non-Abelian  fields living on
them. In other words, the surfaces are endowed with densities of
action and of topological charge.  We argued that such surfaces can be
profiled by 1d defects, or non-Abelian monopoles, whose field is not
spherically symmetric but rather line-like.

For the defects to become dynamical degrees of freedom there should
be a fine tuning which allows infrared and ultraviolet scales to
coexist. Such fine tuning is well understood in the Abelian case,
(see Eq. (\ref{finetuning})). A central point: for the
dual-superconductor mechanism of confinement to be relevant, similar
fine tuning should be realized in the non-Abelian case as well. This
time, however, it is self-tuning between the action and entropy of
the 2d surfaces. No explicit form of the fine tuning is known
theoretically but coexistence of the two scale is seen in the
lattice simulations.

Also,   there should be consistency between dynamics in  four
dimensions, or of the original Yang-Mills fields and dynamics on the
defects world-sheet. Let us conclude with an example of such a
consistency check. For dynamics of a particle (or   1d defect) to be
in fact independent on the ultraviolet scale the action associated
with the 'monopoles' should be approximately
\begin{equation}\label{radiativemass}
L\cdot M_{non-Abelian}~\approx~L\cdot {ln 7\over a}~ ~,
\end{equation}
see Eq. (\ref{finetuning}). And this has been checked on the lattice
\cite{bornyakov}. On the other hand, the radiative mass
(\ref{radiativemass})  is associated now with the non-Abelian fields
which live not in the bulk but on the world-sheet (see discussion
above). Which seems to be also true on the lattice, for references
and review see \cite{zakharov}. The condition (\ref{radiativemass})
replaces the quantization condition for the magnetic field in case
of the non-Abelian monopoles living on the  surfaces.

\section*{Acknowledgments}
We are  thankful to M.N. Chernodub, A.S. Gorsky, F.V. Gubarev,
 Ph. de Forcrand for
useful discussions. This note was worked out mostly   during the time
when both authors were participants to the workshop
``Non-Perturbative Methods in Strongly Coupled Gauge Theories'' at
the Galileo Galilei Institute (Florence). We are thankful to the GGI
for the invitation and hospitality.


\begin{thebibliography}{99}
\bibitem{dirac}
Paul Dirac, {\it "Quantised Singularities in the Electromagnetic Field"},
{\it Proc. Roy. Soc.} {\bf A 133}, (1931) 60.

\bibitem{gukov}
S. Gukov, E. Witten,
 {\it ``Gauge Theory, Ramification, And The Geometric Langlands Program''},
[arXiv:hep-th/0612073];\\
E. Witten,
{\it Fortsch. Phys.}, {\bf 55} (2007) 545.

\bibitem{savit}
R. Savit, { Rev. Mod. Phys.} {\bf 52} (1980) 453.

\bibitem{ambjorn}
A.M. Polyakov, {\it ''Gauge Fields and Strings''},
Harvard Academic Publishers,  (1987);\\
J. Ambjorn, {\it ''Quantization of geometry''}, [arXiv:hep-th/9411179].


\bibitem{reviews}
M.N. Chernodub, F.V. Gubarev, M.I. Polikarpov, A. I. Veselov , {\it
Progr. Theor. Phys. Suppl.}, {\bf 131}. 309 (1998),
[arXiv:hep-lat/9802036];\\
A. Di Giacomo, {\it Progr. Theor. Phys. Suppl.}, {\bf 131} 161 (1998,
[arXiv:hep-lat/9802008]; \\
T. Suzuki, {\it Progr. Theor. Phys. Suppl.}, {\bf 131} 633 (1998).

\bibitem{greensite}
J.~Greensite,
  {\it Prog.\ Part.\ Nucl.\ Phys.},  {\bf 51}    (2003) 1,
  [arXiv:hep-lat/0301023].

\bibitem{polyakov}
A.M. Polyakov,  {\it Phys. Lett.}, {\bf B59} (1975) 82 ;\\
T. Banks, R. Myerson, J. B. Kogut,
{\it Nucl. Phys.}, { \bf B129} (1977) 493;\\
H. Shiba, T. Suzuki,
{\it Phys. Lett.}, {\bf B333} (1994) 461, [arXiv:hep-lat/9404015];\\
A. Di Giacomo, G. Paffuti, {\it Phys. Rev.} {\bf D56} (1997)  6816,
[arXiv:hep-lat/9707003].


\bibitem{stone}
M. Stone, P. R. Thomas,
{\it Phys. Rev. Lett.}, {\bf 41} (1978) 351;\\
M.N. Chernodub, V.I. Zakharov,
 {\it Nucl. Phys.}, {\bf  B669} (2003) 233,
[arXiv:hep-th/0211267].

\bibitem{coleman}
S. R. Coleman, {\it ``The Magnetic Monopole Fifty Years Later''},
published in Erice Subnuclear 1981:21 (QCD161:I65:1981).

\bibitem{ch}
M.N. Chernodub, F.V. Gubarev, M.I. Polikarpov, V. I. Zakharov,
{\it Nucl. Phys.}, {\bf B600}   (2001) 165, [arXiv:hep-th/0010265].

\bibitem{polyakov1}
A.M. Polyakov, {\it Phys. Lett.} {\bf 72B} (1978) 477; B. Svetitsky, L.G. Yaffe,
{\it Nucl. Phys.} {\bf B210} (1982) 423.



\bibitem{pisarski}
R. D. Pisarski, Phys. Rev. D62, 111501 (2000); [arXiv:hep-ph/0101168].


 \bibitem{zakharov}
V.I. Zakharov, {\it Braz. J. Phys.}, {\bf 37}  (2007)  165
[arXiv:hep-ph/0612342]; {\it AIP Conf. Proc.}, {\bf 756} (2005) 182,
[arXiv:hep-ph/0501011].


\bibitem{gorsky}
A. Gorsky, V. Zakharov, {\it Phys. Rev.} {\bf D77} (2008) 045017,
arXiv:0707.1284 [hep-th];\\
A.S. Gorsky, V.I. Zakharov, A. R. Zhitnitsky,
{\it ``On Classification of QCD defects via holography"}'
 [arXiv:0902.1842 [hep-ph]].

\bibitem{goddard}
P. Goddard, J. Nuyts, D. I. Olive,
{\it Nucl. Phys.} {\bf B125} (1977) 1.

\bibitem{sakai}
T. Sakai, Sh. Sugimoto,  {\it Prog. Theor. Phys.} {\bf 113} (2005) 843,
 [arXiv:hep-th/0412141].


\bibitem{thooft}
G. 't Hooft,
  {\it Nucl. Phys.}, {\bf B190} (1981) 455.

\bibitem{lubkin}
E. Lubkin, {\it Ann. Phys.}, {\bf 23}  (1963) 233.

\bibitem{thooft1}
G. 't Hooft,  {\it Nucl. Phys.}, {\bf B153} (1979) 141.

\bibitem{gubarev}
F.V. Gubarev, A.V. Kovalenko, M.I. Polikarpov, S.N. Syritsyn, V.I. Zakharov, {\it Phys. Lett.} {\bf B574} (2003)136,[arXiv:hep-lat/0212003].

\bibitem{kovalenko}
A. V. Kovalenko, S. M. Morozov, M. I. Polikarpov, V. I. Zakharov,
{\it Phys. Lett.} {\bf  B 648} (2007) 383 [arXiv:hep-lat/0512036];\\
R. Hollwieser, M. Faber, J. Greensite, U. M. Heller, S. Olejnik, {\it Phys. Rev.} {\bf D78} (2008) 054508, [arXiv:0805.1846 [hep-lat]].

\bibitem{jubileum}
V.I. Zakharov, {\it``Matter of resolution: From quasiclassics
to fine tuning."},
{\it in} Sense of Beauty in Physics: Miniconference in Honor of
Adriano Di Giacomo on his 70th Birthday, Pisa, Italy, Jan 2006. [arXiv:hep-ph/0602141].

\bibitem{deforcrand}
MILC Collaboration (C. Aubin et al.),
  {\it Nucl. Phys. Proc. Suppl.}
  {\bf 140} (2005) 626,  [arXiv:hep-lat/0410024];\\
 E.M. Ilgenfritz et al.,
{\it Phys. Rev.} {\bf D76} (2007) 034506, [arXiv:0705.0018].

\bibitem{fateev}
A. A. Belavin, V. A. Fateev, A. S. Schwarz and Yu. S. Tyupkin, 
{\it Phys. Lett.} {\bf B83}  (1979);\\
V.A. Fateev, I.V. Frolov, A.S. Shvarts,
   {\it Nucl. Phys.} {\bf B154} (1979)1.

\bibitem{bornyakov}
V.G. Bornyakov {\it et al.},  {\it Phys. Lett.}, {\bf B537} (2002)
291, [arXiv:hep-lat/0103032].

\end{thebibliography}
 \end{document}